\documentclass[letter,onecolumn]{jpsj3}
\usepackage{color}
\usepackage{ulem}

\title{Superconductivity at 38 K in Iron-Based Compound with Platinum-Arsenide Layers
Ca$_{10}$(Pt$_4$As$_8$)(Fe$_{2-x}$Pt$_x$As$_2$)$_5$}

\author{
\name{Satomi \surname{Kakiya}}$^{1,2}$, 
\name{Kazutaka \surname{Kudo}}$^{1,2}$\thanks{E-mail address: kudo@science.okayama-u.ac.jp}, 
\name{Yoshihiro \surname{Nishikubo}}$^{1,2}$, 
\name{Kenta \surname{Oku}}$^{3}$, \\
\name{Eiji \surname{Nishibori}}$^{3}$, 
\name{Hiroshi  \surname{Sawa}}$^{3}$, 
\name{Takahisa \surname{Yamamoto}}$^{4}$, 
\name{Toshio \surname{Nozaka}}$^{5}$, 
\\ and \name{Minoru \surname{Nohara}}$^{1,2}$
}

\inst{
$^{1}$Department of Physics, Okayama University, Okayama 700-8530, Japan \\
$^{2}$JST, Transformative Research-Project on Iron Pnictides (TRIP), Chiyoda, Tokyo 102-0075, Japan \\
$^{3}$Department of Applied Physics, Nagoya University, Nagoya 464-8603, Japan \\
$^{4}$Department of Advanced Materials Science, The University of Tokyo, Kashiwa 277-8561, Japan \\ 
$^{5}$Department of Earth Science, Okayama University, Okayama 700-8530, Japan
}

\abst{
We report superconductivity in novel iron-based compounds Ca$_{10}$(Pt$_n$As$_8$)(Fe$_{2-x}$Pt$_x$As$_2$)$_5$ with $n$ = 3 and 4. 
Both compounds crystallize in triclinic structures (space group $P\bar{1}$), in which Fe$_2$As$_2$ layers alternate with Pt$_n$As$_8$ spacer layers. 
Superconductivity with a transition temperature of 38 K is observed in the $n$ = 4 compound with a Pt content of $x$ $\simeq$ 0.36 in the Fe$_2$As$_2$ layers. The compound with $n$ = 3 exhibits superconductivity at 13 K. }

\kword{Iron-based superconductors, Ca-Fe-Pt-As}

\begin{document}
\maketitle

A number of iron-based superconductors have been identified since the discovery of superconductivity with a transition temperature ($T_c$) of 26 K in the fluorine-doped iron oxypnictide LaFeAsO$_{1-x}$F$_x$\cite{JACS_130_3296_2008}.  
All materials identified so far consist of alternating two-dimensional Fe$_2$As$_2$ layers and ``spacer layers". 
The central issues in realizing higher $T_c$ are to seek novel spacer layers and engineer them to tune the electronic states of Fe$_2$As$_2$ layers in which high-$T_c$ superconductivity emerges.
The crystal structure of iron-based superconductors can be classified according to their spacer layers, which are
(i) alkali or alkali-earth ions, as in LiFeAs~\cite{PhysRevB.78.060505} and BaFe$_2$As$_2$\cite{PhysRevLett.101.107006}, (ii) slabs of rare-earth oxides or alkali-earth fluorides with a fluorite-type structure, as in LaFeAsO~\cite{JACS_130_3296_2008} and CaFeAsF\cite{JACS_130_14428_2008}, and (iii) complex metal oxides with perovskite-type structure or combinations of perovskite-type and rocksalt-type structures, as in Sr$_3$Sc$_2$O$_5$Fe$_2$As$_2$\cite{PhysRevB.79.024516}, Sr$_4$(Sc, Ti)$_3$O$_8$Fe$_2$As$_2$\cite{APEX.3.063102}, Sr$_4$V$_2$O$_6$Fe$_2$As$_2$\cite{PhysRevB.79.220512}, Ca$_4$(Al,Ti)$_2$O$_6$Fe$_2$As$_2$\cite{SST-23-11-115005, shirage:172506}, and their homologous series compounds\cite{SST-22-8-085001, APEX.2.063007, SST-23-4-045001, APEX.3.063102, ogino:072506}. 
All of these spacer layers consist of ionic chemical bonds, thus the spacer layers are insulating in nature.

In this letter, we report the discovery of novel iron-based superconductors with spacer layers made of covalent Pt  arsenides. 
Using X-ray diffraction and chemical analyses, we identified the compounds to be Ca$_{10}$(Pt$_4$As$_8$)(Fe$_{2-x}$Pt$_x$As$_2$)$_5$ (referred to as $\alpha$-phase) and Ca$_{10}$(Pt$_3$As$_8$)(Fe$_{2-x}$Pt$_x$As$_2$)$_5$ (referred to as $\beta$-phase). 
Both compounds crystallize in triclinic structures with the space group $P\bar{1}$ ($\sharp$ 2). 
The $\alpha$-phase exhibits superconductivity at $T_c$ $\simeq$ 38 K, the $\beta$-phase at $T_c$ $\simeq$ 13 K.

Single crystals of Ca$_{10}$(Pt$_n$As$_8$)(Fe$_{2-x}$Pt$_x$As$_2$)$_5$ were grown by heating a mixture of Ca, FeAs, Pt, and As powders. 
The mixture of 0.5 g with a ratio of Ca $:$ Fe $:$ Pt $:$ As  $=$ 23 $:$ 23 $:$ 10 $:$ 44 was placed in an alumina crucible (inner diameter of 6 mm and length of 60 mm) and sealed in an evacuated quartz tube. 
All manipulations were carried out in a glove box filled with argon gas. 
The ampules were heated in two ways. 
Heating at 700 $^\circ$C for 3 h and then at 1000 $^\circ$C for 72 h followed by slow cooling to room temperature yielded an $\alpha$-phase with $T_{\rm c}$ $=$ 38 K, whereas heating at 1100 $^\circ$C and slow cooling to 1050 $^\circ$C for 40 h yielded a $\beta$-phase with $T_{\rm c}$ $=$ 13 K. 
The $\beta$-phase was obtained as a single phase without any secondary phases. 
In contrast, the $\alpha$-phase was always obtained in the form of a mixture of $\alpha$-phase crystals and slag containing nonsuperconducting Ca(Fe$_{1-x}$Pt$_x$)$_2$As$_2$ ($x$ $\simeq$ 0.08) and Ca$_2$As powders. 
We obtained platelike crystals with a typical size of 1 $\times$ 1 $\times$ 0.1 mm$^3$ for both phases. 
The crystals were characterized by synchrotron radiation X-ray diffraction\cite{NIM_467_1045,AIP_1234_887} and electron-probe microanalysis (EPMA). 
Longitudinal and Hall resistivities were measured using a Physical Property Measurement System (Quantum Design). 
Magnetization was measured using a SQUID magnetometer (Quantum Design).

\begin{figure}[t]
\begin{center}
\includegraphics[width=8.5cm]{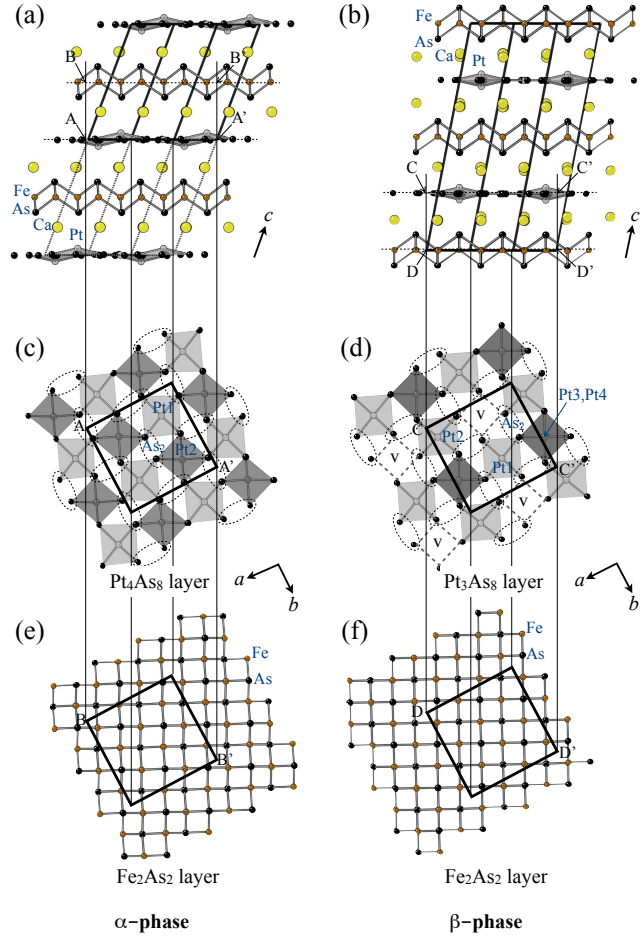}
\caption{
\label{}
(Color online) Crystal structures of Ca$_{10}$(Pt$_4$As$_8$)(Fe$_{2-x}$Pt$_x$As$_2$)$_5$ ($\alpha$-phase) and Ca$_{10}$(Pt$_3$As$_8$)(Fe$_{2-x}$Pt$_x$As$_2$)$_5$ ($\beta$-phase) with triclinic structures (space group $P\bar{1}$ ($\sharp$ 2)). 
Thick solid lines indicate the unit cell. 
(a), (c), and (e) show the schematic overviews of the $\alpha$-phase, Pt$_4$As$_8$ layer, and (Fe$_2$As$_2$)$_5$ layer, respectively. 
The gray and dark-gray hatches in (c) indicate PtAs$_4$ squares with centric Pt1 and off-centered Pt2, respectively. 
The dashed ellipsoids in (c) represent As$_2$ dimers. 
(b), (d), and (f) show the schematic overviews of the $\beta$-phase, Pt$_3$As$_8$ layer, and Fe$_2$As$_2$. 
The gray and dark-gray hatches in (d) indicate PtAs$_4$ squares with centric Pt1 and Pt2 and off-centered Pt3 and Pt4, respectively. 
The occupancy of Pt3 and Pt4 is $\sim$0.5. 
The dashed ellipsoids and `v' in (d) represent As$_2$ dimers and Pt vacancies, respectively. 
}
\end{center}
\end{figure}
\begin{figure}
\begin{center}
\includegraphics[width=6cm]{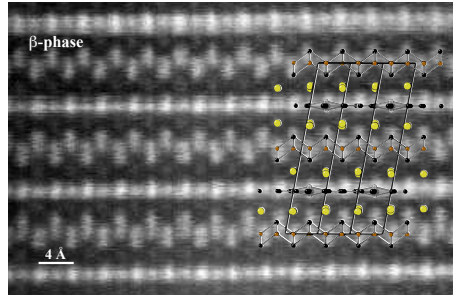}
\caption{\label{}
(Color online) High-resolution TEM image of Ca$_{10}$(Pt$_3$As$_8$)(Fe$_{2-x}$Pt$_x$As$_2$)$_5$ ($\beta$-phase).  
}
\end{center}
\end{figure}
\begin{figure}
\begin{center}
\includegraphics[width=6cm]{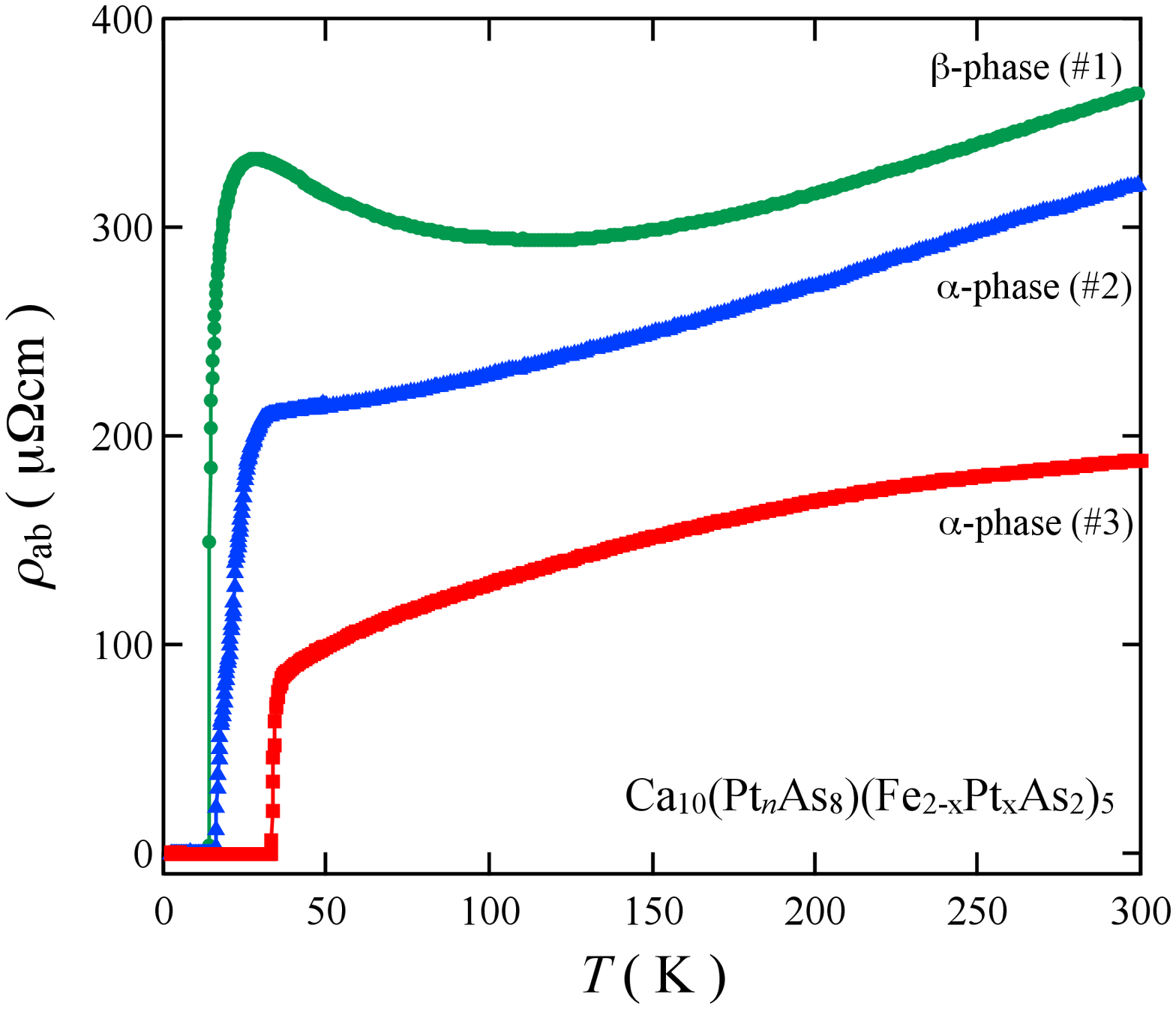}
\caption{\label{}
(Color online) Temperature dependence of electrical resistivity for Ca$_{10}$(Pt$_4$As$_8$)(Fe$_{2-x}$Pt$_x$As$_2$)$_5$ ($\alpha$-phase) and Ca$_{10}$(Pt$_3$As$_8$)(Fe$_{2-x}$Pt$_x$As$_2$)$_5$ ($\beta$-phase). 
}
\end{center}
\end{figure}

Single-crystal structure analysis reveals that the compounds crystallize in triclinic structures with the space group $P\bar{1}$ ($\sharp$ 2). 
The structures consist of alternating stacking of (Fe$_2$As$_2$)$_5$ and Pt$_n$As$_8$ layers ($n$ $=$ 4 for $\alpha$-phase and $n$ $=$ 3 for $\beta$-phase) with five Ca ions between them, as depicted in Figs.~1(a) and 1(b). 
Tiny crystals with dimensions of approximately 50 $\times$ 50 $\times$ 20 $\mu$m$^3$ prevented us from determining the superconducting and normal-state properties of the crystals used for single-crystal structure analysis. 
Thus we performed Rietveld refinement on the well-characterized samples to clarify the relation of crystal structures and chemical compositions with superconducting and normal-state properties. 
The Rietveld refinement reveals the chemical compositions to be Ca$_{10}$(Pt$_4$As$_8$)(Fe$_{2-x}$Pt$_x$As$_2$)$_5$ with $x$ $\simeq$ 0.36 ($\alpha$-phase) and Ca$_{10}$(Pt$_3$As$_8$)(Fe$_{2-x}$Pt$_x$As$_2$)$_5$ with $x$ $\simeq$ 0.16 ($\beta$-phase). 
These chemical formulae yield atomic ratios of Ca $:$ Fe $:$ Pt $:$ As $=$ 23.8 $:$ 19.5 $:$ 13.8 $:$ 42.9 for the $\alpha$-phase and 24.4 $:$ 22.4 $:$ 9.3 $:$ 43.9 for the $\beta$-phase, which are consistent with EPMA results: 22.9 $:$ 23.9 $:$ 9.9 $:$ 43.3 for the $\alpha$-phase and 22.9 $:$ 23.2 $:$ 10.1 $:$ 43.8 for the $\beta$-phase. 
Crystallographic data are summarized in Table I and II.

The most notable feature of the present compounds is the presence of Pt$_n$As$_8$ layers, depicted in Figs.~1(c) and 1(d). 
These platinum arsenide layers are characterized by a distorted square lattice of corner-sharing PtAs$_4$ squares. 
Rotations of the PtAs$_4$ squares result in the formation of As$_2$ dimers, as indicated by the dashed ellipsoids in Figs.~1(c) and 1(d).
Such As$_2$ dimers are observed in pyrite-type structures; the Pt$_4$As$_8$ layers can be derived from the $ab$-plane of pyrite PtAs$_2$. 
The size of the Pt square lattice (with a Pt-Pt distance of approximately 4.4 \AA) is by far larger than the size of the Fe$_2$As$_2$ square lattice (approximately 3.9 \AA~for CaFe$_2$As$_2$). 
This lattice mismatch leads to the formation of the $\sqrt{5} \times \sqrt{5}$ superstructure in the $ab$-plane in the present compounds, as depicted in Figs.~1(e) and 1(f). 
There exist Pt vacancies in the platinum arsenide layers of the $\beta$-phase, as shown in Fig.~1(d), to form the Pt$_3$As$_8$ layers.

The Pt$_ 3$As$_8$ layers alternating with Fe$_2$As$_2$ layers were directly confirmed for the $\beta$-phase, as shown in Fig.~2, by high-angle annular dark-field scanning transmission electron microscopy (HAADEF-STEM, JEM-2100F, JEOL Co. Ltd.) with an aberration corrector (CEOS GmbH). 
In this image, intensity is approximately proportional to $Z^2$, where $Z$ is the atomic number; thus the brightest spots correspond to Pt, and the faintest spots correspond to Ca. 
Ca is located between the Fe$_2$As$_2$ and Pt$_3$As$_8$ layers, consistent with the X-ray refinements.

Figure 3 shows the electrical resistivity of $\alpha$- and $\beta$-phases of different batches as a function of temperature. 
For sample $\sharp$3 with the $\alpha$-type structure, the resistivity shows metallic behavior. 
The resistivity starts to decrease sharply at 37 K and becomes negligibly small below 32.7 K. 
The 10\%$-$90\% transition width is approximately 2 K. 
This result suggests that the sample becomes superconducting below 32.7 K. 
The resistivity of $\sharp$1 with the $\beta$-type structure exhibits semiconducting behavior below about 110 K,  followed by a decrease below 27 K. 
Zero resistivity was observed below 13.5 K. 
The resistivity of $\sharp$2 with the $\alpha$-type structure shows an intermediate behavior between those observed in $\sharp$1 and $\sharp$3.

\begin{figure}
\begin{center}
\includegraphics[width=6cm]{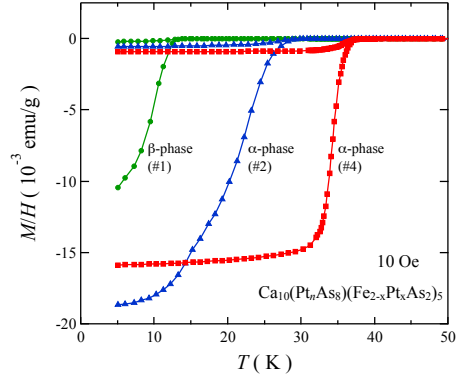}
\caption{\label{}
(Color online) Temperature dependence of dc magnetization $M$ measured in a magnetic field $H$ of 10 Oe for Ca$_{10}$(Pt$_4$As$_8$)(Fe$_{2-x}$Pt$_x$As$_2$)$_5$ ($\alpha$-phase) and Ca$_{10}$(Pt$_3$As$_8$)(Fe$_{2-x}$Pt$_x$As$_2$)$_5$ ($\beta$-phase). 
}
\end{center}
\end{figure}

\begin{figure}[t]
\begin{center}
\includegraphics[width=6cm]{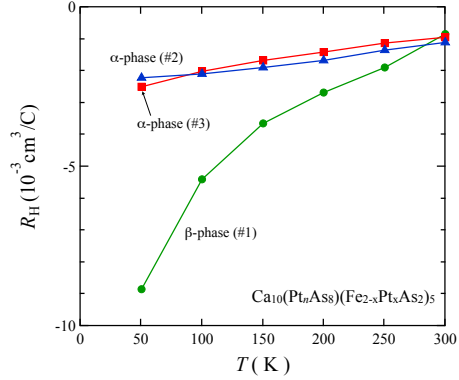}
\caption{\label{}
(Color online) Temperature dependence of Hall coefficient for Ca$_{10}$(Pt$_4$As$_8$)(Fe$_{2-x}$Pt$_x$As$_2$)$_5$ ($\alpha$-phase) and Ca$_{10}$(Pt$_3$As$_8$)(Fe$_{2-x}$Pt$_x$As$_2$)$_5$ ($\beta$-phase).
}
\end{center}
\end{figure}

Evidence of bulk superconductivity can be seen in the temperature dependence of magnetization $M$ shown in Fig.~4.  
All samples exhibited a diamagnetic behavior, a hallmark of superconductivity. 
The onset $T_c$'s were estimated to be 38 K, 29 K, and 13 K for $\sharp$4, $\sharp$2, and $\sharp$1, respectively. 
The shielding volume fraction at 5 K without demagnetizing-field corrections were 139, 163, and 82\% for $\sharp$4, $\sharp$2, and $\sharp$1, respectively. 
Thus the highest $T_c$ of 38 K was obtained in the $\alpha$-phase ($\sharp$4).

%\begin{table}
\begin{table}[p]
\caption{
Crystallographic parameters of $\alpha$-phase. 
$M$ site represents Fe$_{1-x/2}$Pt$_{x/2}$ with $x$ = 0.36(4). 
Chemical formula is Ca$_{10}$Fe$_{8.2}$Pt$_{4.8}$As$_{18}$. 
The occupancy is fixed at 1 in all atomic sites. 
$R_{wp}$ $=$ 6.39\% and $RI$ $=$ 7.67\%. 
}
\begin{tabular}{cccc}
\hline
\hline
\multicolumn{4}{c}{The $\alpha$-phase}\\
\hline
\multicolumn{4}{l}{Triclinic with space group $P\bar{1}$ ($\sharp$ 2).} \\
\multicolumn{4}{l}{$a$ $=$ 8.719(1) \AA, $b$ $=$ 8.727(1) \AA, $c$ = 11.161(1) \AA.} \\
\multicolumn{4}{l}{$\alpha$ $=$ 99.04(2)$^\circ$, $\beta$ $=$ 108.21(2)$^\circ$, $\gamma$ $=$ 90.0(2)$^\circ$.} \\
\hline
\multicolumn{4}{c}{atomic positions}\\
site & $x/a$ & $y/b$ & $z/c$ \\%
Ca01 &   0.338(1) &  0.2818(6) &  0.2404(2) \\%
Ca02 &   0.135(1) & 0.6868(6)  & 0.2404(2) \\%
Ca03 &   0.542(1) &  0.8785(6) &  0.2404(2) \\%
Ca04 &   0.744(1) &  0.4850(6) &  0.2404(2) \\%
Ca05 &  0.935(1) &  0.0794(6)  & 0.2404(2) \\%
$M$01 & 0.645(2) &  0.453(4) &  0.500(1) \\%
$M$02 & 0.846(2) & 0.053(4)  &  0.500(1) \\%
$M$03 & 0.246(2) &  0.253(4) &  0.500(1) \\%
$M$04 & 0.446(2) &  0.853(4) &  0.500(1) \\%
$M$05 & 0.044(2) &  0.653(4) &  0.500(1) \\%
Pt01  &   0.750(2)  & 0.250(2)    &  0.000(1) \\%
Pt02  &    0.234(1)  &  0.235(2)  &   0.9421(5)  \\%
As01 &  0.4854(6)  &  0.6155(9)  &  0.36805(7) \\%
As02 &  0.2849(6)  &  0.0175(9)  &  0.36805(7) \\%
As03 &  0.6842(6)  &  0.2167(9)  &  0.36805(7) \\%
As04 &  0.0861(6)  &  0.4168(9)  &  0.36805(7) \\%
As05 &   0.8857(6)  &  0.8171(9) &   0.36805(7) \\%
As06 &   0.848(3)    & -0.496(3)   &  1.004(3)  \\%
As07 &  -0.345(3)   & -0.025(3)    &  1.004(3)  \\%
As08 &  -0.005(3)   & -0.142(3)    &  1.004(3)  \\%
As09 &   0.504(3)   & -0.372(3)    &  1.004(3)  \\%
\hline
\hline
\end{tabular}
\end{table}

%\begin{table}
\begin{table}[p]
\caption{
Crystallographic parameters of $\beta$-phase. 
$M$ site represents Fe$_{1-x/2}$Pt$_{x/2}$ with $x$ = 0.16(1). 
Chemical formula is Ca$_{10}$Fe$_{9.20}$Pt$_{3.80}$As$_{18}$. 
The occupancy is fixed at 1 in all atomic sites except for Pt03 and Pt04. 
The occupancy is determined to be 0.447(3) at Pt03 and Pt04.
$R_{wp}$ $=$ 4.82\% and $RI$ $=$ 7.77\%. 
}
\begingroup
\renewcommand{\arraystretch}{0.7}
\begin{tabular}{cccc}
\hline
\hline
\multicolumn{4}{c}{The $\beta$-phase}\\
\hline
\multicolumn{4}{l}{Triclinic with space group $P\bar{1}$ ($\sharp$ 2).} \\
\multicolumn{4}{l}{$a$ $=$ 8.795(3) \AA, $b$ $=$8.789(3) \AA, $c$ = 21.008(7) \AA.} \\
\multicolumn{4}{l}{$\alpha$ $=$ 94.82(8)$^\circ$, $\beta$ $=$ 99.62(9)$^\circ$, $\gamma$ $=$ 89.99(3)$^\circ$.} \\
\hline
\multicolumn{4}{c}{atomic positions}\\
site & $x/a$ & $y/b$ & $z/c$ \\%
Ca01 &  0.0005(3) & 1.1278(3) & 0.3667(1) \\%   
Ca02 & -0.2915(3) & 1.4761(3) & 1.1335(1) \\%   
Ca03 & -0.1978(3) & 1.2793(3) & 0.6326(1) \\%   
Ca04 & -0.3940(3) & 1.6750(3) & 0.6327(1) \\%  
Ca05 & 0.1976(3) & 1.4779(3) & 0.6330(1) \\%  
Ca06 & -0.2997(3) & 1.7221(3) & 0.8674(1) \\%   
Ca07 & 0.0928(3) & 0.9186(3) & 0.8669(1) \\%   
Ca08 &  -0.5095(3) & 1.1200(3) & 0.8519(1) \\%   
Ca09 & -0.1035(3) & 1.3267(3) & 0.8680(1) \\%  
Ca10 & -0.5906(3) & 1.0794(3) & 0.6480(1) \\%  
$M$01 & -0.1522(2) & 0.5506(2) & 0.9999(6) \\%   
$M$02 & -0.0503(2) & 0.8516(2) & 1.0014(6) \\%   
$M$03 & -0.2525(2) & 0.2492(2) & 0.9998(7) \\%   
$M$04 &  -0.3518(2) & 0.9503(2) & 0.9999(8) \\%   
$M$05 & -0.4506(2) & 0.6508(2) & 0.9991(7) \\% 
$M$06 &  0.1497(2) & 0.9501(2) & 0.4992(7) \\%   
$M$07 & -0.3514(2) & 1.4492(2) & 0.49866(6) \\%   
$M$08 & -0.4501(2) & 1.1490(2) & 0.50007(7) \\%   
$M$09 &  0.2506(2) & 1.2478(2) & 0.5002(7) \\%   
$M$10 & -0.0506(2) & 1.3481(2) & 0.50013(6) \\%   
Pt01 &  0.45028(6) & 0.59961(5) & 0.74990(2) \\% 
Pt02 & -0.04976(6) & 0.10073(5) & 0.75011(2) \\%  
Pt03 & -0.0389(1) & 0.6055(1) & 0.77790(4) \\%  
Pt04 & -0.0615(1) & 0.5943(1) & 0.72210(4) \\%  
As01 & 0.2187(2) & 0.9996(1) & 0.75004(5) \\%   
As02 & -0.4513(2) & 0.8670(1) & 0.75008(5) \\%   
As03 & 0.3249(2) & 1.0356(2) & 0.43135(6) \\%  
As04 & -0.2920(2) & 1.4827(2) & 0.75010(7) \\%  
As05 & -0.5236(2) & 1.3607(2) & 0.56856(6) \\%   
As06 & -0.1245(2) & 1.5642(2) & 0.56678(6) \\%   
As07 & -0.1813(2) & 1.0362(2) & 0.93135(7) \\%   
As08 & 0.0670(2) & 1.3569(2) & 0.74993(8) \\%  
As09 & -0.3803(2) & 1.4343(2) & 0.93125(7) \\%  
As10 &  0.0237(2) & 1.6376(2) & 0.93316(7) \\%   
As11 & 0.0785(2) & 1.1619(2) & 0.56872(6) \\%   
As12 & -0.2255(2) & 1.7656(2) & 1.06898(7) \\%   
As13 & -0.1693(2) & 0.8453(2) & 0.74997(7) \\%   
As14 & -0.2780(2) & 1.2332(2) & 0.43126(6) \\%   
As15 & 0.3500(2) & 1.3384(1) & 0.74993(6) \\%  
As16 & -0.5776(2) & 1.8381(2) & 0.93099(7) \\%  
As17 & 0.1940(2) & 0.7191(2) & 0.74993(7) \\%  
As18 & -0.3118(2) & 1.1976(1) & 0.75022(6) \\%  
\hline
\hline
\end{tabular}
\endgroup
\end{table}

The Hall coefficient $R_H$, shown in Fig.~5, is negative for all samples, suggesting that the major carriers are electrons in both $\alpha$- and $\beta$-phases. 
The strong temperature dependence of $R_H$ observed in the $\beta$-phase ($\sharp$1), as well as the semiconducting resistivity at low temperatures, is reminiscent of underdoped materials. 
Indeed, analogous behaviors are observed in the underdoped Ba(Fe$_{1-x}$Co$_x$)$_2$As$_2$ with $x$ = 0.05\cite{JPSJ.78.123702}. 
$R_H$ for the $\alpha$-phase ($\sharp$2 and 3) exhibits a weak but distinct temperature dependence, which again is analogous to those reported in optimally doped Ba(Fe$_{1-x}$Co$_x$)$_2$As$_2$ with $x$ = 0.09 \cite{JPSJ.78.123702}. 
The Pt content of $x/2$ $\simeq$ 0.18 for optimally doped Ca$_{10}$(Pt$_4$As$_8$)(Fe$_{2-x}$Pt$_x$As$_2$)$_5$ is by far larger than the Co content for optimally doped Ba(Fe$_{1-x}$Co$_x$)$_2$As$_2$ ($x$ = 0.09), suggesting that Pt causes the doping of fewer electrons than Co, the most studied dopant for iron-based materials. This is consistent with the absence of superconductivity in Ca(Fe$_{1-x}$Pt$_x$)$_2$As$_2$ up to the solubility limit of $x$ $\simeq$ 0.08.

Formal electron counts of As$_2$ dimers and isolated As are [As$_2$]$^{4-}$ and As$^{3-}$, respectively. 
Since all As atoms form dimers in Pt$_n$As$_8$ layers and all As atoms are isolated in Fe$_2$As$_2$ layers, we expect a formal electron count of Fe$^{2+}$ and Pt$^{2+}$ for the $\beta$-phase, Ca$_{10}$(Pt$_3$As$_8$)(Fe$_{2}$As$_2$)$_5$. Thus,  Ca$_{10}$(Pt$_3$As$_8$)(Fe$_{2}$As$_2$)$_5$ can be viewed as a parent compound.

The novel spacer layers of the present materials, namely, Pt$_n$As$_8$ ($n$ $=$ 3 and 4) layers, will give us a unique opportunity to further engineer iron-based materials to optimize superconductivity. 
First, the present materials with an interlayer distance between Fe$_2$As$_2$ of $d$ $\simeq$ 10 \AA \ fill the materials gap between SmFeAsO with $d$ $\simeq$ 8.6 \AA \ and Sr$_4$V$_2$O$_6$Fe$_2$As$_2$ with $d$ $\simeq$ 15.67 \AA.  
Smaller $d$ than that of SmFeAsO tends to lead to lower $T_c$\cite{ogino:072506, Lee}. 
Larger $d$ than that of Sr$_4$V$_2$O$_6$Fe$_2$As$_2$ also seems to lead to lower $T_c$\cite{ogino:072506}.  
We naively expect that an optimal distance for higher $T_c$ may exist in the $d$ range of the present materials. 
Secondly, the nature of chemical bonds of the spacer layers seems to be quite different between the present materials and the previous iron-based superconductors. 
The spacer layers of the previous materials, alkali/alkali-earth ions or complex oxides/fluorides, are electrically inert because of the ionic chemical bonds. 
Thus no atomic orbitals of the spacer layers mix with the Fe $3d$ orbitals of the superconducting Fe$_2$As$_2$ layers. 
In contrast, the chemical bonds of the present spacers layers, a platinum arsenide, are covalent in nature; thus we expect that Pt orbitals may contribute to charge transport and possibly to superconductivity. 
Indeed, most platinum arsenides are metallic and even superconductivity has been observed in SrPt$_2$As$_2$\cite{JPSJ.79.123710} and SrPtAs\cite{JPSJ.80.055002}. 
Further investigations will be invaluable to determine whether novel electronic states are realized in the present materials.

Recently, Ni {\it et al.}\cite{arXiv.1106.2111} reported superconductivity and crystal structures on Ca$_{10}$(Pt$_4$As$_8$)(Fe$_2$As$_2$)$_5$ and Ca$_{10}$(Pt$_3$As$_8$)(Fe$_2$As$_2$)$_5$, based on our previous report\cite{NS2-2011}. 
The $\alpha$-phase and $\beta$-phase in the present study most likely correspond to 10-4-8 and 10-3-8 phases reported by Ni {\it et al.}\cite{arXiv.1106.2111}. 
Higher $T_{\rm c}$ was observed in the present study, likely due to the difference in Pt contents in the Fe$_2$As$_2$ layers.

In summary, we reported that the quaternary iron-based materials Ca$_{10}$(Pt$_4$As$_8$)(Fe$_{2-x}$Pt$_x$As$_2$)$_5$ and Ca$_{10}$(Pt$_3$As$_8$)(Fe$_{2-x}$Pt$_x$As$_2$)$_5$ exhibit superconductivity at 38 K and 13 K, respectively. 
Both materials consist of novel platinum-arsenide spacer layers, namely, Pt$_n$As$_8$ layers. 
This finding demonstrated that a variety of spacer layers are possible in iron-based superconductors. 
Engineering of this novel spacer layer will lead to higher superconducting transition temperatures in this class of materials.

We thank C.-H. Lee for valuable discussion and M. Takasuga, S. Nakano, M. Danura, and S. Pyon for their technical assistance. 
Part of this work was performed at the Advanced Science Research Center, Okayama University, Japan. 
This work was partially supported by KAKENHI from JSPS and MEXT, Japan. 
The synchrotron radiation experiments performed at BL02B1 and BL02B2 of SPring-8 were supported by the Japan Synchrotron Radiation Research Institute (JASRI) (Proposal No. 2010A, B0083, 0084).

{\it Note added in proof} -- We noticed a paper by L\"{o}hnert {\it et al.} [arXiv: 1107.5320] reporting similar results with $T_{\rm c}$ up to 35 K, based on our previous report\cite{NS2-2011}.

\end{document}